# Simulation Studies of Nanomagnet-Based Logic Architecture

*David B. Carlton\*, Nathan C. Emley\*, Eduard Tuchfeld, and Jeffrey Bokor*

Department of Electrical Engineering and Computer Science,

University of California, Berkeley, CA 94720

\* These authors contributed equally to this work

Corresponding Author. Email: dcarlton@eecs.berkeley.edu. Telephone: 310.614.1109

ABSTRACT. We report a simulation study on interacting ensembles of Co nanomagnets that can perform basic logic operations and propagate logic signals, where the state variable is the magnetization direction. Dipole field coupling between individual nanomagnets drives the logic functionality of the ensemble and coordinated arrangements of the nanomagnets allow for the logic signal to propagate in a predictable way. Problems with the integrity of the logic signal arising from instabilities in the constituent magnetizations are solved by introducing a biaxial anisotropy term to the Gibbs magnetic free energy of each nanomagnet. The enhanced stability allows for more complex components of a logic architecture capable of random combinatorial logic, including horizontal wires, vertical wires, junctions, fanout nodes, and a novel universal logic gate. Our simulations define the focus of scaling trends in nanomagnet-based logic and provide estimates of the energy dissipation and time per nanomagnet reversal.

One of the most significant challenges facing scaling trends in conventional Si CMOS technology is the amount of power dissipation per unit area. Consequently, interest is very high in logic technologies and architectures whose functionality does not rely on passing an electron current. One technology demonstrating basic logic operations while passing *no* electric current is based on ensembles of lithographically patterned, interacting nanomagnets[1,2]. For nanomagnets placed at small separation (i.e. sub-20nm), effects of the magnetostatic fringe field, or dipole field, on one nanomagnet from its neighbor can be sufficiently strong as to perturb its magnetization direction. Coordinated arrangements of these interacting nanomagnets allow for predictable outcomes for each magnetization direction, given controlled inputs. Thus, it is possible to design logic gates that are driven by the dipolar coupling and employ magnetization direction, rather than electric charge, as the state variable.

Here we report the results of a detailed micromagnetic simulation study of key components of a complete nanomagnet-based logic architecture. Several new concepts in nanomagnetic logic are introduced and validated by our simulations. We demonstrate that by engineering the constituent nanomagnets with a magnetocrystalline biaxial anisotropy, the enhanced stability of the ensembles allows the construction of more complex elements, including vertical and horizontal wires, junctions, fan-out nodes, and a novel universal logic gate. To highlight scaling trends, we simulate logic propagation in horizontal wires as a function of nanomagnet dimensions and biaxial anisotropy constant. The important values of switching time and energy dissipation per nanomagnet volume are estimated. These simulations are an important step towards realizing a low power, nanomagnet-based logic architecture.

Each nanomagnet is fabricated with elongated shape-defined uniaxial anisotropy, with the long axis of all nanomagnets oriented parallel to one another with edge-to-edge spacing ~20 nm. Both logic operation and signal propagation are initialized by first aligning all the magnetizations along their magnetically hard (lithographically short) axes by application of a global external magnetic field $H_{ext}$. Once $H_{ext}$ is removed, the logic signal is propagated along the wire by supplying, as input, a perturbative magnetic field at one end of the wire, say, from the output nanomagnet of a logic gate. This field cants the magnetization of the first nanomagnet in the wire, pushing it out of its metastable hard axis alignment. The magnetization of the second nanomagnet then feels a push from the dipole field of the first and is likewise pushed off its hard axis, and so on down the wire. The resulting behavior is analogous to a toppling line of dominoes, with each magnetization relaxing from hard to easy axis alignment in a cascading fashion.

Propagation of the cascade in a horizontal wire predictably leads to sequentially antiparallel magnetizations[3] (logic 1, logic 0, logic 1, etc.), which can be thought of as a string of NOT operations. Success of this cascade is contingent upon the magnetizations remaining hard axis-aligned for the duration of the cascade transient. However, previous work has shown that moderately long logic cascades typically fail[4]. The predominant failure originates in the tenuous hard axis metastability of the individual nanomagnets. To date, experimental[1-4] and computational[5,6] research has considered nanomagnets either with uniaxial anisotropy or with no anisotropy. The Gibbs magnetic free energy or magnetization energy for the uniaxial nanomagnets is $U(\theta) = K_u \cos^2(\theta)$, where $K_u$ is the uniaxial anisotropy constant (here $K_u$ is shape-defined) and $\theta$ is the in-plane magnetization angle relative to the hard axis. Although hard axis alignment is an unstable configuration for an isolated nanomagnet, many nanomagnets assembled in a horizontal wire boost hard axis stability through nearest-neighbor dipole fields that favor parallel alignment when all the moments point along their hard axes. However, such a configuration relies on all of the moments to retain their direction until affected by the logic cascade. If one nanomagnet were to fall out of hard axis alignment, its dipole field would upset the direction of its neighbors, leading to aberrant cascade nucleations. Effects from temperature fluctuations, stray fields, or lithographic irregularities can destabilize hard axis alignment. For longer propagation times (i.e. longer wires) the probability of an instability to nucleate increases, significantly weakening the prospect for nanomagnetic interconnects.

We introduce the concept of enhancing hard axis stability by adding a biaxial anisotropy term to the net magnetization energy of each nanomagnet. Unlike the nearest-neighbor dipole field boost in stability mentioned above, which is a product of the total nanomagnet ensemble and beneficial only to the extent that all moments remain hard axis-aligned, this new stability is intrinsic to the constituent nanomagnets. We introduce a magnetostatic energy of the form:

(1) $\quad U(\theta) = K_u\cos^2(\theta) + \tfrac{1}{4}K_1\sin^2(2\theta)$

where $K_1$ is the biaxial anisotropy constant and the biaxial and uniaxial anisotropy axes are coincident. In this single domain picture, the biaxial term reduces or even inverts the curvature of $U(\theta)$ at $\theta = 0°$, thereby enhancing hard axis stability. For our simulations we employ the publicly available OOMMF simulator [7], use a cell size of $(5\text{ nm})^3$, saturation magnetization (for Co) $M_s = 10^6$ A/m, and exchange stiffness $A = 1.3 \times 10^{-11}$ J/m. $U(\theta)$ plots are generated for a 100 nm × 50 nm rectangle (with 10 nm × 10 nm squares removed from each corner) by first saturating the magnetization at an angle $\theta$, then removing the saturation field and returning the total energy 0.1 ps later. For all nanomagnet dimensions studied in this work, reversal is observed to be domain wall-dominated, likely due to the highly non-uniform dipole fields that drive the signal propagation, but the $U(\theta)$ plots for nanomagnets coerced to be single domain illustrate the stabilizing effect of the biaxial anisotropy. We plot $U(\theta)$ for $K_1 = 0$, 30, and 60 kJ/m$^3$ in Figure 1(a). Fitting to each curve reliably yields $K_u = 34$ kJ/m$^3$. Although the majority of our simulations are carried out for zero temperature (for the sake of computational convenience) we highlight the benefit of finite biaxial anisotropy by simulating a nanomagnet wire at $T = 300$K and show the comparison in Figure 1(b), where the wire with no biaxial anisotropy fails after only four nanomagnets. Temperature dependence in micromagnetic simulations is implemented in the standard manner by augmenting the standard Landau-Lifshitz-Gilbert equation with a Langevin, stochastically fluctuating temperature dependent field[8].

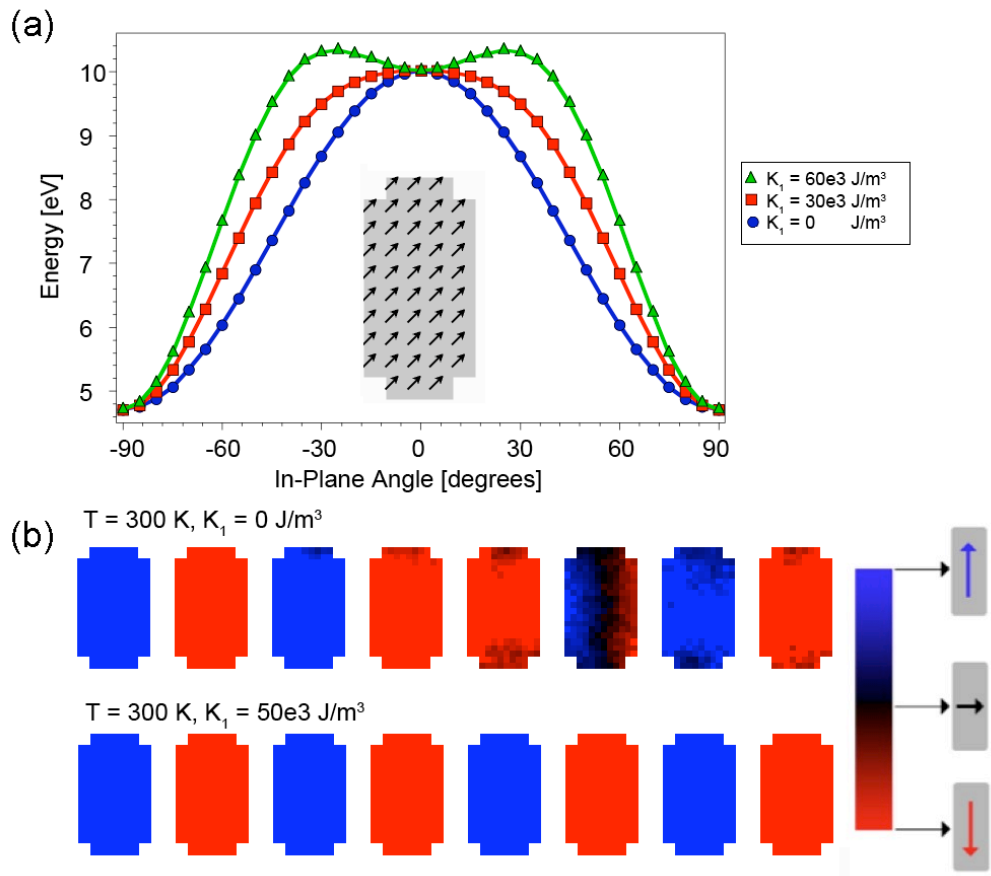

**Figure 1.** (a) Magnetic energy $U(\theta)$, from OOMMF simulations of the nanomagnet shown in the inset (at $\theta = 45°$), for $K_1 = 0$, 30, and 60 kJ/m$^3$. The biaxial anisotropy term alters the curvature of the $U(\theta)$ plot near $\theta = 0°$, making that point (hard axis) more stable. (b) Simulated wires of eight nanomagnets at $T = 300$K with $K_1 = 0$ and $K_1 = 50$ kJ/m$^3$.

To quantitatively explore the impact of $K_1$ we simulate a horizontal wire of fifteen rectangular Co nanomagnets and plot the success of the cascade propagation as a function of nanomagnet length, width, and $K_1$, with nanomagnet thickness $t = 5$ nm and separation $d = 20$ nm. Total length of the simulations was 3 ns. Shown in Figure 2, each plot consists of three distinct regions, identified by their x-y plane coloring: *light gray*, where logic propagation fails due to incorrect switching of nanomagnets whose aspect ratios (length:width) are too large, rendering them unstable; *dark gray*, where logic does not propagate within the 3 ns simulation time because the nanomagnets are too stable (aspect ratios are too small); *white*, where logic propagation is successful within 3 ns. The wedge of parameter space with successful logic propagation (white region) demonstrates a working range of nanomagnet aspect ratios and shows that scaling nanomagnet dimensions into the sub-50 nm dimension regime is possible.

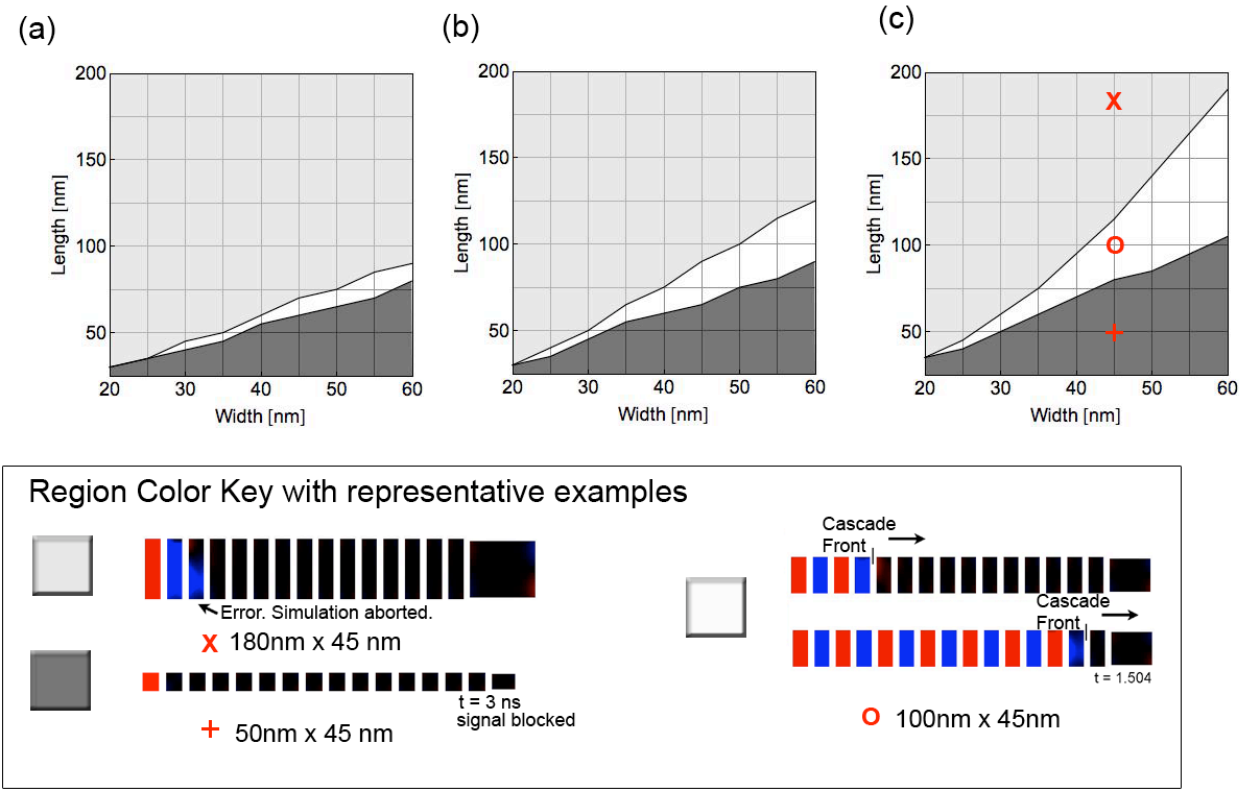

**Figure 2.** Plots of whether or not the signal propagated successfully as a function of nanomagnet length, width, and amount of biaxial anisotropy. In (a), (b), and (c) the nanomagnets have biaxial anisotropy constant $K_1$ = 30, 40, and 50 kJ/m$^3$ respectively. White wedge-shaped areas indicate regions of parameter space for which the signal propagated correctly with all nanomagnets flipping in the correct order within the 3 ns simulation time. Grey areas are where signal propagation was unsuccessful. Failure mechanisms are explained in the text. False positives, where an instability along the wire led

fortuitously to proper alignment of the magnetizations, are accounted for in the analysis.

Vertical wires, where the nanomagnets are stacked with their short axes abutting one another, in addition to horizontal wires are necessary for realizing a general nanomagnetic interconnect scheme. Logic propagation along a vertical wire is also initiated by aligning all the moments along their hard axes, where they must remain until the logic cascade arrives and reorients them to all align parallel, up or down. Unlike horizontal wires, nearest neighbor dipole fields in vertical wires oppose the magnetization direction and thereby hinder hard axis stability. Even for values of $K_1$ that stabilize horizontal wires, the added biaxial anisotropy is insufficient to ensure successful logic propagation in vertical wires. A solution to this nearest-neighbor-induced instability is to place square "stabilizer" nanomagnets to the left and right of each nanomagnet in the vertical wire. With their magnetization directions set by $H_{ext}$, dipole fields from the stabilizers compensate for those from the nearest neighbors, and because of their square shape, shape-induced anisotropy is minimal. The biaxial anisotropy, in conjunction with their large moment, makes them relatively impervious to stray dipole fields from the nanomagnetic wires. The necessity for these stabilizers in vertical wires is discussed in greater detail in the supplementary online material. Signal fanout and arbitrary routing of the logic is also necessary, which requires junctions of vertical and horizontal wires where the logic signal must branch out and form copies of itself. Such branching and fanout is demonstrated in the supplementary online material.

Dynamics of the cascade are controlled by the hard-to-easy axis relaxation rate of individual nanomagnets and the strength of the dipolar coupling between them. As the cascade transit is magnon-like, logic propagation times will be considerably longer than electrical signals in conventional electronic interconnects. Propagation times from the simulations shown in Figure 2 are 1 – 2 ns for fifteen nanomagnets, or ~100 ps per nanomagnet. Energy dissipation, which comes from viscous damping of the magnetization and exchange energy costs associated with domain wall dynamics, is examined in our simulations for nanomagnets with dimensions 100 nm × 50 nm (10nm squares removed from the corners), $t = 5$ nm, $d = 20$ nm, with $K_1 = 50$ kJ/m$^3$. Energy dissipation per nanomagnet reversal is 2.5 eV. Details of the simulation are provided in the supplementary online material. A similar simulation for 50 nm × 30 nm nanomagnets (all other parameters unchanged) yields an energy dissipation per reversal of 1.0 eV. The difference between the ratio of dissipation energies (2.5) and the ratio of nanomagnet volumes (4.2) for the two simulations likely rests in the domain wall dominated nature of the reversal. This is supported by the fact that the 5.3 eV single domain nanomagnet energy barrier calculated from $U(\theta)$ (Fig. 1(a)) for $\theta = 90°$ to $0°$ magnetization rotation is much larger than that for domain wall reversal. We point out that our choice of initial conditions, where all nanomagnets begin uniformly magnetized to the right, is equivalent to H_ext being instantaneously removed at time = 0. It is understood that this will lead to a much larger amount of energy dissipated than if H_ext were removed adiabatically, at a timescale much larger than the magnetization relaxation time[9,10].

The 3-input majority logic (MLG) gate experimentally demonstrated by Imre et al.[1] is the current state-of-the-art for nanomagnet-based logic. Three magnetic wires serve as input to the gate by surrounding and terminating at a central nanomagnet whose magnetization is set along its hard axis. When the logic signals arrive at the central nanomagnet, the combined effects of the dipole fields from the three adjacent nanomagnets determine the final orientation, up or down, along the easy axis of the central nanomagnet. Each of the three inputs thus supplies a vote for the direction of the central nanomagnet, and whichever direction receives the majority of votes wins.

One issue with this approach is that if the inputs do not arrive at precisely the same time the gate cannot be made to behave predictably. Consider the example shown in Figure 3(a). If two inputs arrive simultaneously with opposite votes, then the central moment is unaffected and the output must be determined by the vote of the third input. However, in the opposite case where a single input arrives first, its one vote must not be sufficient to flip the central nanomagnet. Thus, we have two contradictory scenarios with the distinction between them resting only in the arrival timing of the three inputs. The problem of this so-called race condition would be eliminated from the MLG if all inputs are forced to be synchronous. This condition may be achievable in a cellular automata information processing architecture. However, for complex combinatorial logic circuits this is virtually impossible. Consequently, design of a different logic gate that does not suffer from such race conditions is desirable. We introduce a modified, 2-input universal logic gate that solves the problem of asynchronous inputs by breaking up the logic operation into two complimentary sub-gates.

The first sub-gate is a modified MLG, where the strength of inputs 1 and 3 are designed to have weaker coupling to the central nanomagnet and so can only affect an output when their votes agree. Weaker coupling is achieved by spacing the two inputs slightly further from the central nanomagnet. This removes the need for inputs 1 and 3 to arrive synchronously as the weaker coupling effectively gives each input only half a vote. Input 2 must then be given majority rule through stronger coupling as it alone must cause the central nanomagnet to flip if inputs 1 and 3 disagree. The race conditions for all three inputs could then be removed if input 2 is engineered to arrive last and act as a clock signal to supply a potentially deciding vote in the case of a tie between inputs 1 and 3. Due to the asymmetry in voting power of the input islands, we refer to this modified majority logic gate as the 3-input "Dictator" gate (D-gate) to emphasize the role played by input 2 in determining the output. While the D-gate solves the race condition problem for inputs 1 and 3, input 2 must be engineered to not arrive first. This is accomplished by the second complimentary sub-gate.

The second sub-gate, which completes our modified 2-input universal logic gate, has a truth table as shown in Figure 3(b). The gate acts as a logical AND gate with the exception that when it should output 1 it outputs nothing at all. We refer to it, then, as the "Lazy AND" gate. As will be shown below, in order for this gate to work it must have the ability to discriminate and block different incoming signals. We achieve this by including nanomagnets that can only switch in one direction — only down in our implementation. Although this unidirectional bias can be achieved using antiferromagnetic pinning layers, doing so would lead to considerable fabrication complexity. Instead, this unidirectional anisotropy is engineered by placing stabilizer nanomagnets at opposite corners of those nanomagnets in the incoming signal wire, as shown in Figure 3(b). The combined dipole field from these two new stabilizers is preferentially down, does not switch the nanomagnet down, but does prevent the nanomagnet from switching up. We refer to this as a "magnetic diode" because a signal is propagated only if it possesses the correct polarity when it arrives at the unidirectional nanomagnet. The complete Lazy AND gate consists of two of these magnetic diodes that surround and strongly couple to the central nanomagnet so that either vote alone is sufficient to output a result. If both inputs to the magnetic diodes are 'up' (logic 1), both are blocked and the central nanomagnet is unaffected. In all other cases one of the signals will be a 'down' (logic 0) and will be passed through, as indicated by the truth table in Figure 3(b).

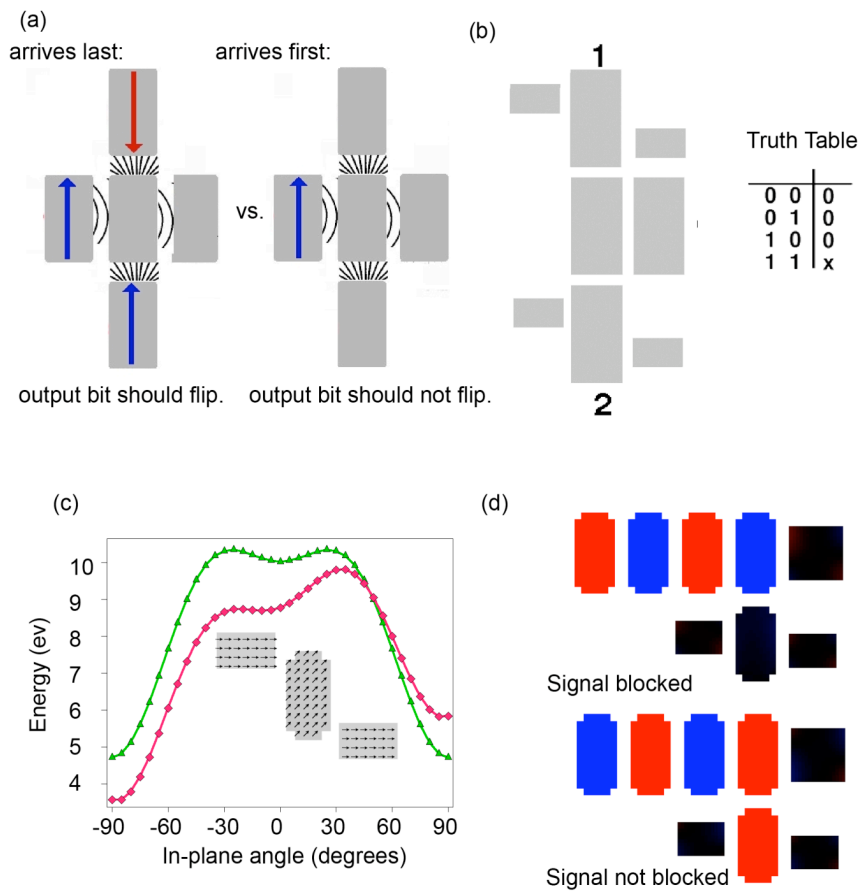

**Figure 3.** (a) Shows how the arrival of the left input in two different scenarios can lead to contradictory requirements for functionality. (b) The structure and truth table of the Lazy AND gate. (c) The magnetic

energy $U(\theta)$, of a nanomagnet with uniaxial + biaxial anisotropy and a nanomagnet biased by unidirectional stabilizers (magnetic diode) that is shown in the inset. (d) OOMMF simulation showing how the magnetic diode preferentially blocks the signal.

The complete, unclocked 2-input universal logic gate is created by connecting the inputs to the Lazy AND gate as well as to inputs 1 and 3 of the Dictator gate and then connecting the output of the Lazy AND to input 2 of the Dictator gate. This arrangement is shown in Figure 4. In each case where the Lazy AND provides an output (01, 10, 00), that signal is passed to the D-gate's strongly coupled input (input 2) and the output is immediately determined. Because the other inputs to the D-gate will either be cancelling (01, 10) or in agreement with input 2 (00) the arrival of the inputs 1 and 3 is not important. In the only case where the Lazy AND gate does not produce an output due to both signals being blocked (11), the output will be determined by inputs 1 and 3 of the D-gate whose two 'up' votes are enough to produce the desired output. Micromagnetic simulations for this gate are shown for 2 different inputs at 2 times in Figure 4.

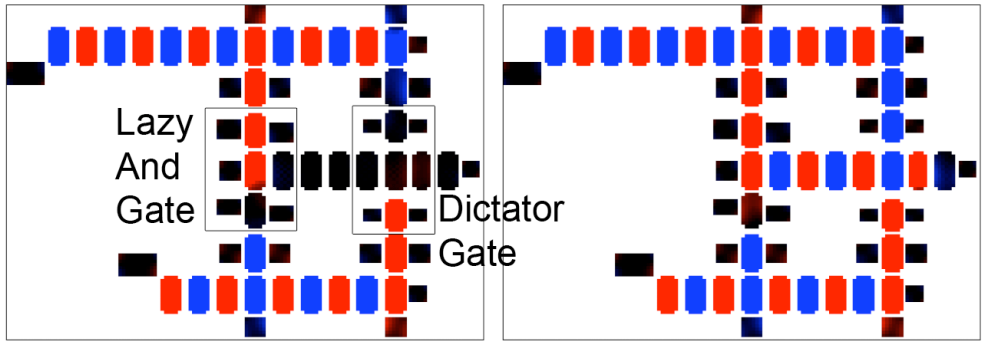
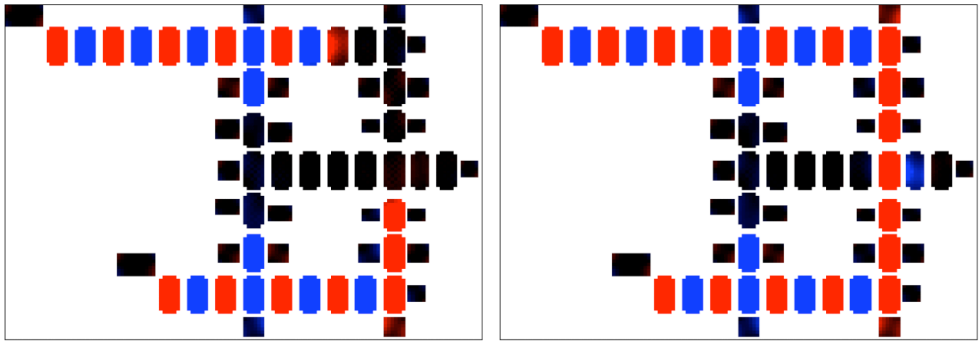

**Figure 4.** (a) NAND gate showing how it is constructed from a Lazy AND a Dictator gate. Operation of the NAND gate at 2 different times with input 01,10. The first input is blocked, but the second input passes through the Lazy AND and determines the output. Input 00 (not shown) operates in a similar manner with both inputs being passed through by the Lazy AND. (b) Operation of the NAND gate at 2 different times with input 11. Both inputs are blocked by the Lazy AND. The output is determined the by 2 inputs going into the D-gate. Full movies of the micromagnetic simulations for all input combinations are available in the supplementary material.

The truth table of this gate can correspond to that of a standard NAND gate. However, since the NOT operation in this architecture can be implemented by adding/subtracting an extra nanomagnet to/from any input or output signal, this same structure can easily be used with no alteration to create a NOR, AND, or OR gate. The logical operation that this structure performs is thus only dependent upon the number of nanomagnets in the wires connecting it with the next gate.

In conclusion, we have employed the publicly available OOMMF micromagnetic simulator package to explore and develop new ideas in a nanomagnet-based logic architecture. A biaxial anisotropy term added to the magnetic energy $U(\theta)$ of each nanomagnet enhances their hard axis stability sufficiently to ensure successful logic propagation in more complex structures of a complete logic architecture, including wires, junctions, fanout nodes, and a novel universal logic gate. Our simulations show how these ensembles may scale down nanomagnet length scales to sub-50 nm. These simulations are an important step towards implementing a working logic architecture based entirely on patterned nanomagnetic elements.

**Acknowledgements:** This work was supported in part by the Western Institute of Nanotechnology.

**Supporting Information Available:** Movies of additional micromagnetic simulations. This material is available free of charge via the Internet at http://pubs.acs.org.

**References**

[1] A. Imre, G. Csaba, L. Ji, A. Orlov, G. H. Bernstein, and W. Porod, Science 311, 205 (2006).

[2] R. P. Cowburn and M. E. Welland, Science 287, 1466 (2000).

[3] R. P. Cowburn, Phys. Rev. B 65, 092409 (2002).

[4] A. Imre, G. Csaba, G. H. Bernstein, W. Porod, and V. Metlushko, Superlatt. Microstruct. 34, 513 (2003).

[5] G. Csaba, A. Imre, G. H. Bernstein, W. Porod, and V. Metlushko, IEEE Trans. Nanotech. 1, 209 (2002).

[6] M. C. B. Parish and M. Forshaw, Appl. Phys. Lett. 83, 2046 (2003).

[7] M. J. Donahue and D. G. Porter, OOMMF User's Guide, Version 1.0 Interagency Report NISTIR 6376, National Institute of Standards and Technology, Gaithersburg, 1999.

[8] J. Xiao, A. Zangwill, and M. Stiles, Phys. Rev. B 72, 014446 (2005).

[9] G. Csaba, P. Lugli, W. Porod, IEEE Trans. Nanotech, p. 346 (2004)

[10] B. Behin-Aein, S. Salahuddin, S. Datta, http://arxiv.org/abs/0804.1389 (2008) [Preprint]